\documentclass[aps,prl,twocolumn,superscriptaddress,showpacs]{revtex4}

\usepackage{amsmath}
\usepackage{amsfonts}
\usepackage{amssymb}
\usepackage{graphicx}
 \usepackage[bf]{subfigure}

\newcommand{\DD}{2D}
\newcommand{\TD}{3D}

\newcommand{\WTMM}{WTMM}
\newcommand{\WTMMM}{WTMMM}
\newcommand{\WT}{WT}
\newcommand{\gra}[1]{{\mathbf #1}}
\newcommand{\ron}[1]{{\mathcal #1}}
\newcommand{\bpsi}{\boldsymbol{\psi}}
\newcommand{\gb}{\gra{b}}
\newcommand{\Mpsi}{\ron{M}_{\bpsi}}
\newcommand{\Tpsirho}{{\mathbf T}_{\bpsi,\rho}}
\newcommand{\carreBlanc}{\protect\scalebox{0.6}{\ensuremath{\square}}}
\newcommand{\rondBlanc}{\ensuremath{\circ}}
\newcommand{\carreNoir}{\protect\scalebox{0.6}{\ensuremath{\blacksquare}}}
\newcommand{\rondNoir}{\ensuremath{\bullet}}

\begin{document}

\title{Generalizing the wavelet-based multifractal formalism to vector-valued random fields: application to turbulent velocity and vorticity 3D numerical data}

\author{Pierre Kestener}
\affiliation{CEA-Saclay, DSM/DAPNIA/SEDI, 91191 Gif-sur-Yvette, France}
\affiliation{Laboratoire de Physique, Ecole Normale Sup\'erieure de Lyon, 46 all\'ee d'Italie, 69364 Lyon c\'edex 07, France}
\author{Alain Arneodo}
\affiliation{Laboratoire de Physique, Ecole Normale Sup\'erieure de Lyon, 46 all\'ee d'Italie, 69364 Lyon c\'edex 07, France}

\date{\today}

\begin{abstract}
We use singular value decomposition techniques to generalize the
wavelet transform modulus maxima method to the multifractal
analysis of vector-valued random fields. The method is calibrated on 
synthetic multifractal 2D vector measures and monofractal 3D
fractional Brownian vector fields. We report the results of some
application to the velocity and vorticity fields issued from 3D
isotropic turbulence simulations. This study reveals the existence of
an intimate relationship between the singularity spectra of these two
vector fields which are found significantly more intermittent than
previously estimated from longitudinal and transverse velocity
increment statistics.
\end{abstract}


\maketitle

The multifractal formalism was introduced in the context of
fully-developed turbulence data analysis and modeling to account for
the experimental observation of some deviation to Kolmogorov theory
(K41) of homogenous and isotropic turbulence~\cite{bFri95}.
The predictions of various multiplicative cascade models, including
the weighted curdling (binomial) model proposed by
Mandelbrot~\cite{aMan74b}, were tested using box-counting (BC) estimates of
the so-called $f(\alpha)$ singularity spectrum of the dissipation
field~\cite{aMen91}. Alternatively, the intermittent nature of the
velocity fluctuations were investigated via the computation of the
$D(h)$ singularity spectrum using the structure function (SF)
method~\cite{aPar85}.
Unfortunately, both types of studies suffered from severe
insufficiencies. On the one hand, they were mostly limited by one
point probe measurements to the analysis of one (longitudinal)
velocity component and to some 1D surrogate
approximation of the dissipation~\cite{aAur92}.
On the other hand, both the BC and SF
methodologies have intrinsic limitations and fail to fully
characterize the corresponding singularity spectrum since only the
strongest singularities are a priori amenable to these
techniques~\cite{aMuz91aMuz94aArn95}. 
In the early nineties, a wavelet-based statistical approach was
proposed as a unified multifractal description of singular measures
and multi-affine functions~\cite{aMuz91aMuz94aArn95}.
Applications of the so-called {\it wavelet transform modulus maxima}
(WTMM) method have already provided insight into a wide variety of
problems, e.g., fully developed turbulence, econophysics, meteorology,
physiology and DNA sequences~\cite{bArn95,bDis02}.
Later on, the WTMM method was generalized to 2D for multifractal
analysis of rough surfaces~\cite{aArnDec00b}, with very promising
results in the context of the geophysical study of the intermittent
nature of satellite images of the cloud
structure~\cite{aArn99caArnDec00c,bArn02} and the medical assist in
the diagnosis in digitized mammograms~\cite{bArn02,akes01}.
Recently the WTMM method has been further extended to 3D analysis
and applied to dissipation and enstrophy 3D numerical data issue from
isotropic turbulence direct numerical simulations
(DNS)~\cite{aKes03_prl,tKes03}.
Thus far, the multifractal description has been
mainly devoted to scalar measures and functions. In the spirit of a
preliminary theoretical study of self-similar vector-valued measures
by Falconer and O'Neil~\cite{aFal96}, our objective here is to
generalize the WTMM method to vector-valued random fields with the
specific goal to achieve a comparative 3D vectorial multifractal
analysis of DNS velocity and vorticity fields.

Let us note $\mathbf{V}(\mathbf{x}=(x_1,x_2,..,x_d))$, a 
vector field with  
square integrable scalar components $V_j(\mathbf{x})$, $j=1,2,..,d$.
Along the line of the 3D WTMM
method~\cite{aKes03_prl,tKes03}, let us define $d$ wavelets
$\psi_i(\mathbf{x}) = \partial \phi(\mathbf{x})/\partial x_i$ 
for $i=1,2,..,d$ respectively, where $\phi(\mathbf{x})$ is a
scalar smoothing function well localized around $|\mathbf{x}|=0$.
The wavelet transform (WT) of $\mathbf{V}$ at point $\mathbf{b}$ and
scale $a$ is the following tensor~\cite{tKes03}:
\begin{equation}
{\mathbb T}_{\bpsi} [\mathbf{V}] ({\mathbf b},a) = 
\begin{pmatrix}
T_{\psi_1}[V_1] & T_{\psi_1}[V_2] & ... & T_{\psi_1}[V_d]\\
T_{\psi_2}[V_1] & T_{\psi_2}[V_2] & ... & T_{\psi_2}[V_d]\\
\vdots & \vdots & & \vdots\\
T_{\psi_d}[V_1] & T_{\psi_d}[V_2] & ... & T_{\psi_d}[V_d]\\
\end{pmatrix},
\label{eq1}
\end{equation}
where 
\begin{equation}
T_{\psi_i}[V_j](\mathbf{b},a) = a^{-d} \int d^d\mathbf{r}\; \psi_i
\bigl( a^{-1} (\mathbf{r} - \mathbf{b}) \bigr) V_j(\mathbf{r}).
\label{eq2}
\end{equation}
In order to characterize the local H\"older
regularity of $\mathbf{V}$,
one needs to find
the direction that locally corresponds to the maximum amplitude
variation of $\mathbf{V}$. This can be obtained from the {\it singular
value decomposition} (SVD)~\cite{bGol89} of the matrix $(T_{\psi_i}[V_j])$
(Eq.~(\ref{eq1})):
\begin{equation}
{\mathbb T}_{\bpsi} [\mathbf{V}] = \mathbb{G} \Sigma \mathbb{H}^T\, ,
\label{eq3}
\end{equation}
where $\mathbb{G}$ and $\mathbb{H}$ are orthogonal matrices
($\mathbb{G}^T\mathbb{G}=\mathbb{H}^T\mathbb{H}=\mathbb{I}_d$) and
$\Sigma=diag (\sigma_1,\sigma_2,..,\sigma_d)$ with $\sigma_i \geq 0$,
for $1\leq i\leq d$.
The columns of $\mathbb{G}$ and $\mathbb{H}$ are referred to as the
left and right singular vectors, and the singular 
values of ${\mathbb T}_{\bpsi} [\mathbf{V}]$ 
are the non-negative square roots $\sigma_i$ of the
$d$ eigenvalues of ${\mathbb T}_{\bpsi} [\mathbf{V}]^T{\mathbb
  T}_{\bpsi} [\mathbf{V}]$.
Note that this decomposition is unique, up to some permutation
of the $\sigma_i$'s.
The direction of the largest amplitude variation of $\mathbf{V}$, at
point $\mathbf{b}$ and scale $a$, is
thus given by the eigenvector $\mathbf{G}_{\rho}(\mathbf{b},a)$
associated to the spectral radius $\rho(\mathbf{b},a)=\max_j
\sigma_j(\mathbf{b},a)$. One is thus led to the analysis of the vector
field ${\mathbf T}_{\bpsi,\rho} [\mathbf{V}](\mathbf{b},a) =
\rho(\mathbf{b},a) \mathbf{G}_{\rho}(\mathbf{b},a)$. 
Following the WTMM analysis of scalar
fields~\cite{aArnDec00b,aKes03_prl,tKes03}, let us define, at a given
scale $a$, the WTMM as the position $\mathbf{b}$ where the modulus
$\Mpsi [\mathbf{V}](\gb,a) = |\Tpsirho [\mathbf{V}](\gb,a)| =
\rho(\gb,a)$ is locally maximum along the direction of
$\mathbf{G}_{\rho}(\mathbf{b},a)$.
These WTMM lie on connected $(d-1)$ hypersurfaces called {\it maxima
  hypersurfaces} (see Figs \ref{fig2}b and \ref{fig2}c). In theory, at
each scale $a$, one only needs to record the position of the local
maxima of $\Mpsi$ (WTMMM) along the maxima hypersurfaces together with
the value of $\Mpsi[\mathbf{V}]$ and the direction of
$\mathbf{G}_{\rho}$. These WTMMM are disposed along connected curves
across scales called {\it maxima
  lines}
living in a $(d+1)$ space
$(\mathbf{x},a)$. The WT skeleton is then defined as the set of
maxima lines that converge to the $(x_1,x_2,..,x_d)$ hyperplane in the
limit $a\rightarrow 0^+$ (see Fig.~\ref{fig2}d). The local H\"older
regularity of $\mathbf{V}$ is estimated from the power-law behavior 
$\Mpsi[\mathbf{V}] \bigl( \ron{L}_{{\mathbf r_0}}(a) \bigr) \sim
a^{h(\gra{r}_0)}$ along the maxima line $\ron{L}_{{\mathbf r_0}}(a)$
pointing to the point ${\mathbf r_0}$ in the limit $a\rightarrow 0^+$,
provided the H\"older exponent $h({\mathbf r_0})$ be smaller than the
number $n_{\bpsi}$ of zero moments of the analyzing wavelet
$\bpsi$~\cite{remark}. As for scalar
fields~\cite{aMuz91aMuz94aArn95,aArnDec00b,aKes03_prl}, the
tensorial WTMM method consists in defining the partition functions:
\begin{equation}
{\mathcal Z}(q,a)=\sum_{{\mathcal L}\in {\mathcal L}(a)} \left (
  \ron{M}_{\bpsi}[\mathbf{V}]({\mathbf r},a)\right)^q\; \sim \;
  a^{\tau(q)}\, ,
\label{eq4}
\end{equation}
where $q \in \mathbb{R}$ and ${\mathcal L}(a)$ is the set of maxima
lines that exist at scale $a$ in the WT skeleton. Then by Legendre
transforming $\tau(q)$, one gets the singularity spectrum $D(h)=\min_q
(qh-\tau(q))$, defined as the Hausdorff dimension of the set
of points $\mathbf{r}$ where $h(\mathbf{r})=h$. Alternatively, one can
compute the mean quantities:
\begin{equation}
\begin{aligned}
h(q,a)=&
\sum_{{\mathcal L}\in{\mathcal L}(a)} \ln \left| 
\ron{M}_{\bpsi}[\mathbf{V}]({\mathbf r},a) \right| \; W_{\bpsi}[\mathbf{V}](q,{\mathcal L}, a)\; ,
\raisebox{-0.5cm}{(5)}
\label{eq56}
\\
D(q,a)=&
\sum_{{\mathcal L}\in{\mathcal L}(a)} W_{\bpsi}[\mathbf{V}](q,{\mathcal L}, a) \; \ln
\bigl( W_{\bpsi}[\mathbf{V}](q,{\mathcal L}, a) \bigr) \; ,
\notag
\end{aligned}
\end{equation}
where 
$W_{\bpsi}[\mathbf{V}](q,{\mathcal L}, a)=\bigl(\ron{M}_{\bpsi}[\mathbf{V}]({\mathbf r},a) \bigr)^q/{\mathcal Z}(q,a)$ 
is a Boltzmann weight computed from the WT skeleton. From the scaling
behavior of these quantities, one can extract $h(q) =\lim_{a\rightarrow 0^+} h(q,a)/\ln a$ and $D(q) =\lim_{a\rightarrow 0^+} D(q,a)/\ln a$
and therefore the $D(h)$ spectrum.

\begin{figure}
  \centering
  \subfigure[]{\includegraphics[scale=0.23]{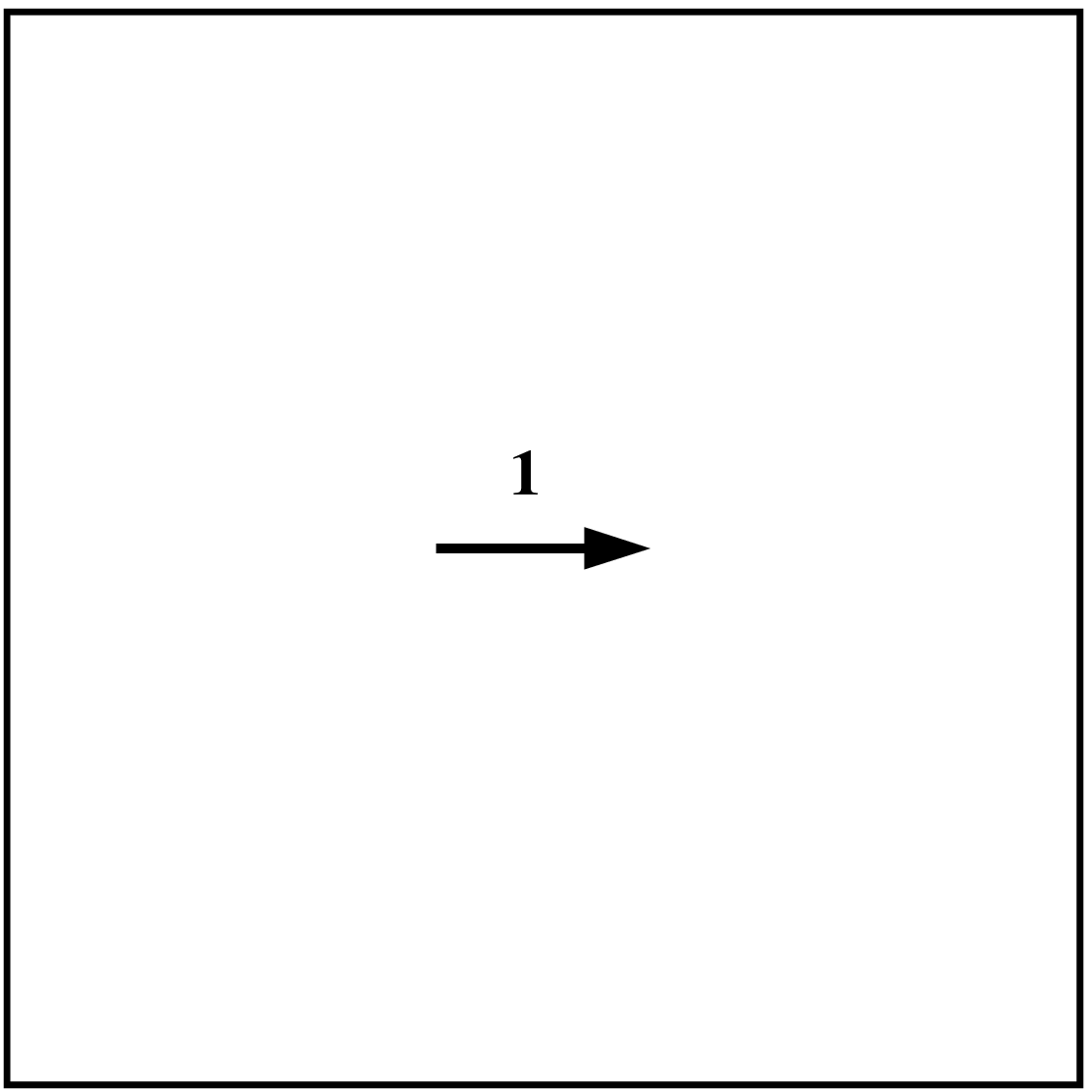}}
  \subfigure[]{\includegraphics[scale=0.23]{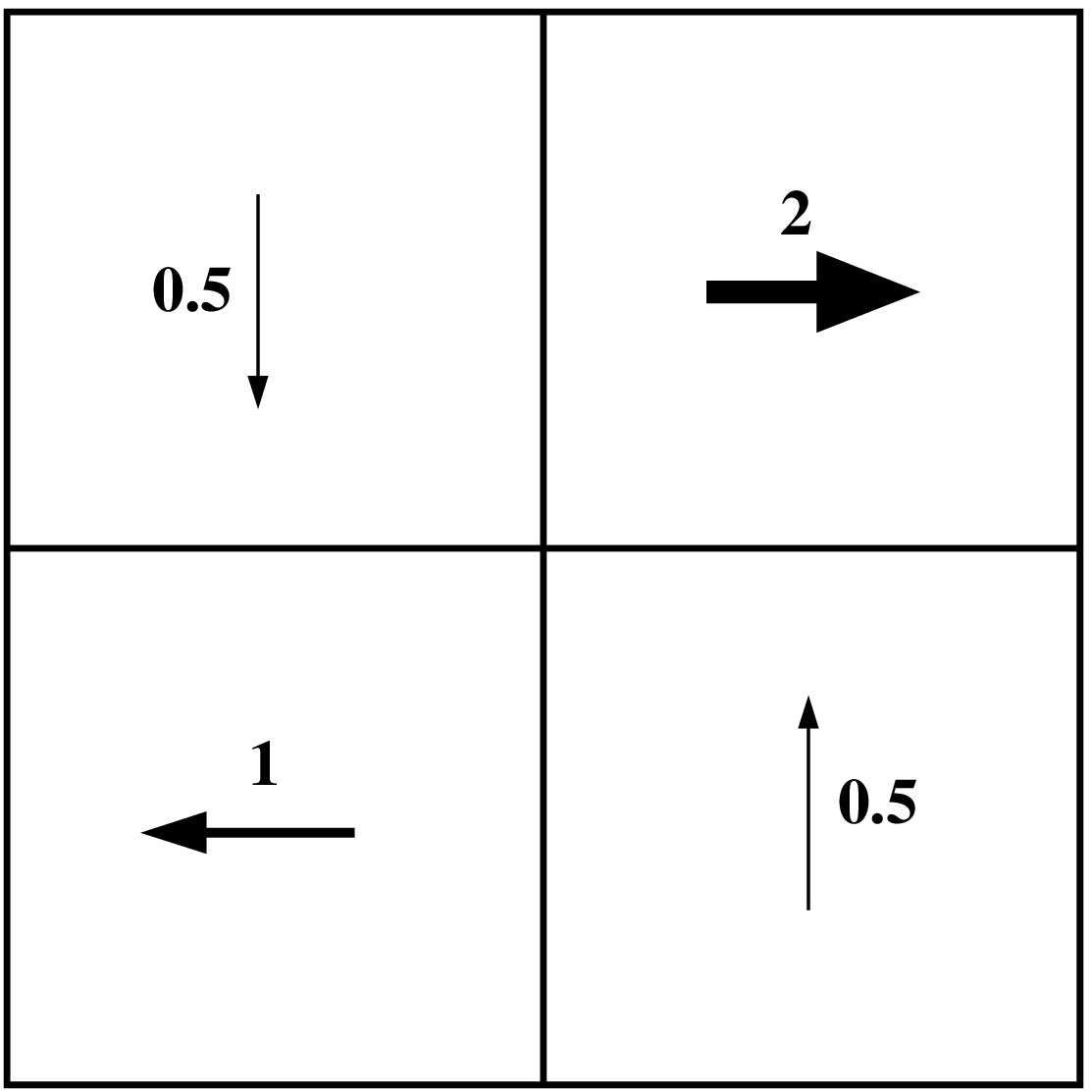}}
  \subfigure[]{\includegraphics[scale=0.23]{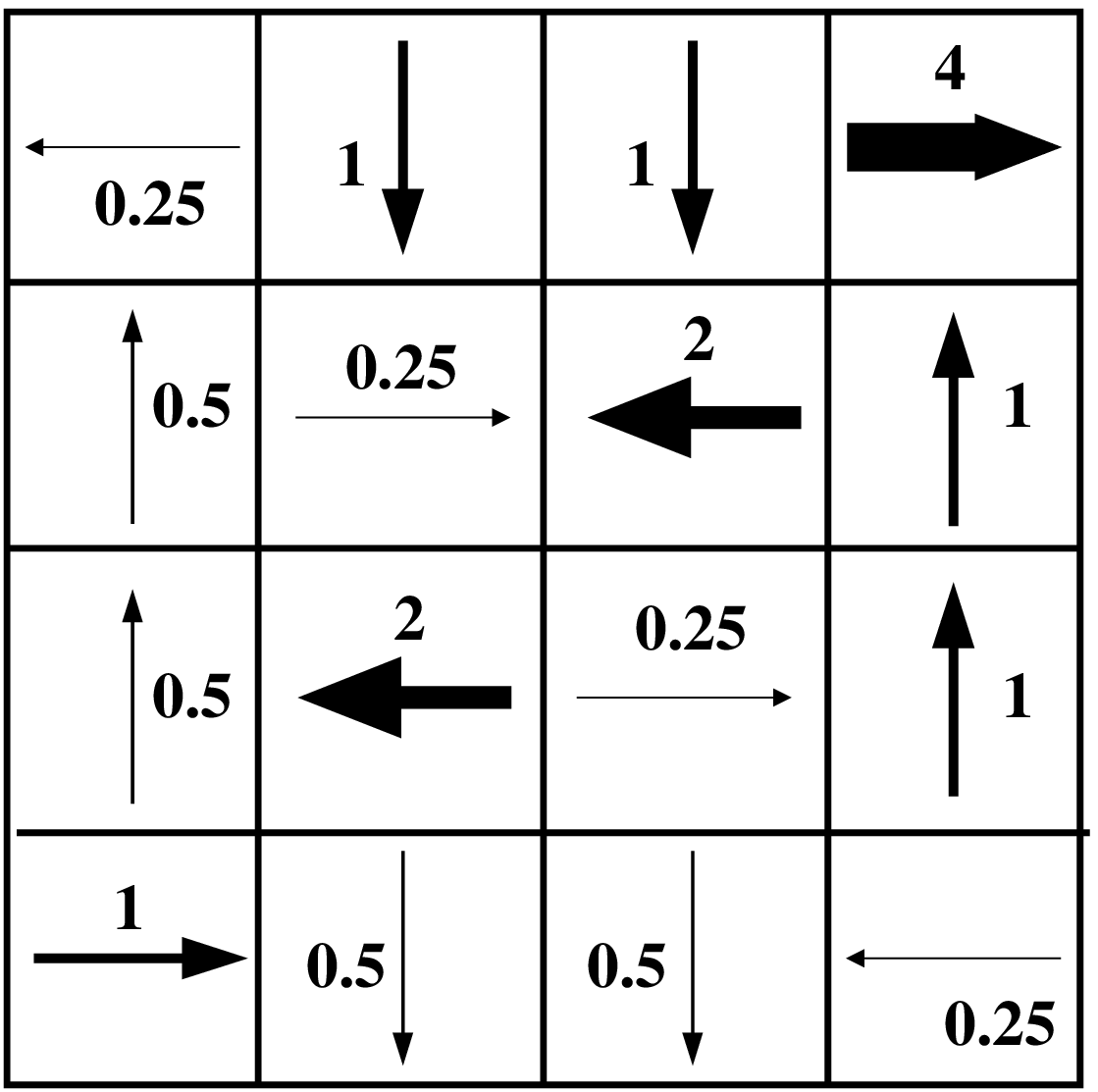}}
  \caption{First construction steps of a 
  singular vector-valued measure supported by the unit square. The
  norm of the four similitude $S_i$ are $p_1=p_4=1/2$, $p_2=2$ and
  $p_3=1$~\cite{aFal96}.} 
  \label{fig1}
\end{figure}

\begin{figure}
  \begin{minipage}[c]{.45\linewidth}
    \includegraphics[width=4.2cm]{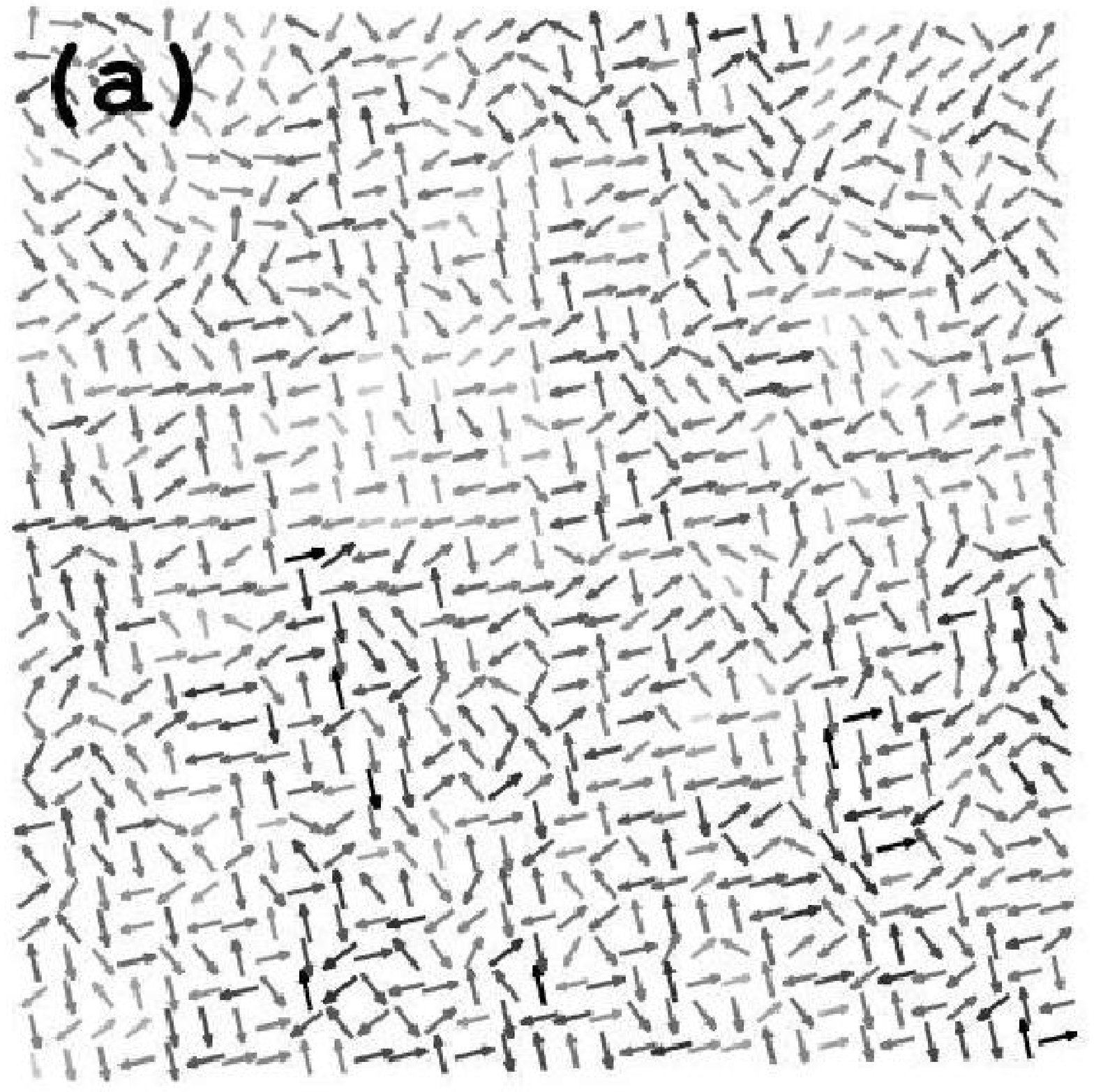}
  \end{minipage}
  \hfill
  \begin{minipage}[c]{.45\linewidth}
    \includegraphics[width=4.2cm]{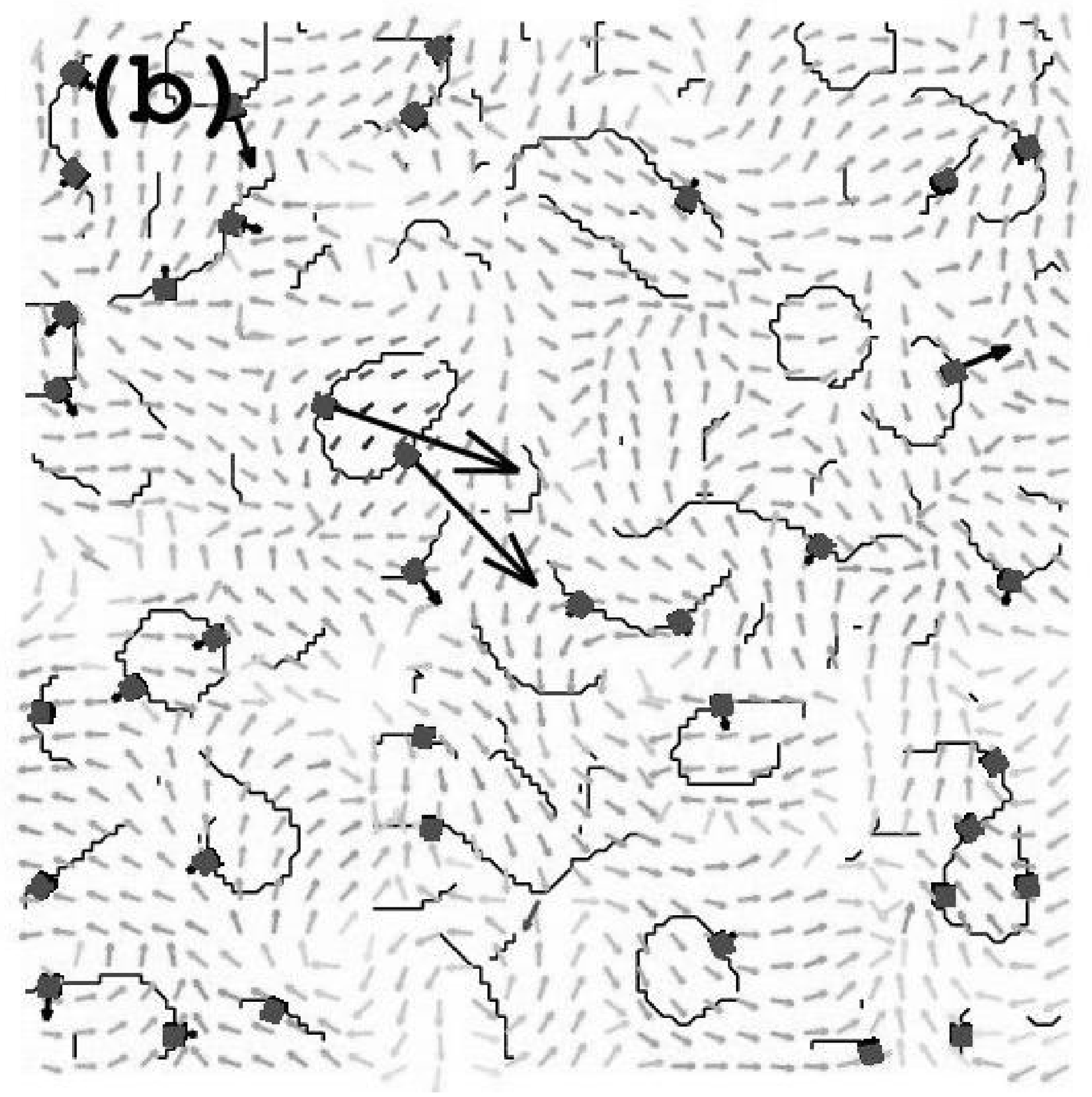}
  \end{minipage}
  \begin{minipage}[c]{.45\linewidth}
    \includegraphics[width=4.2cm]{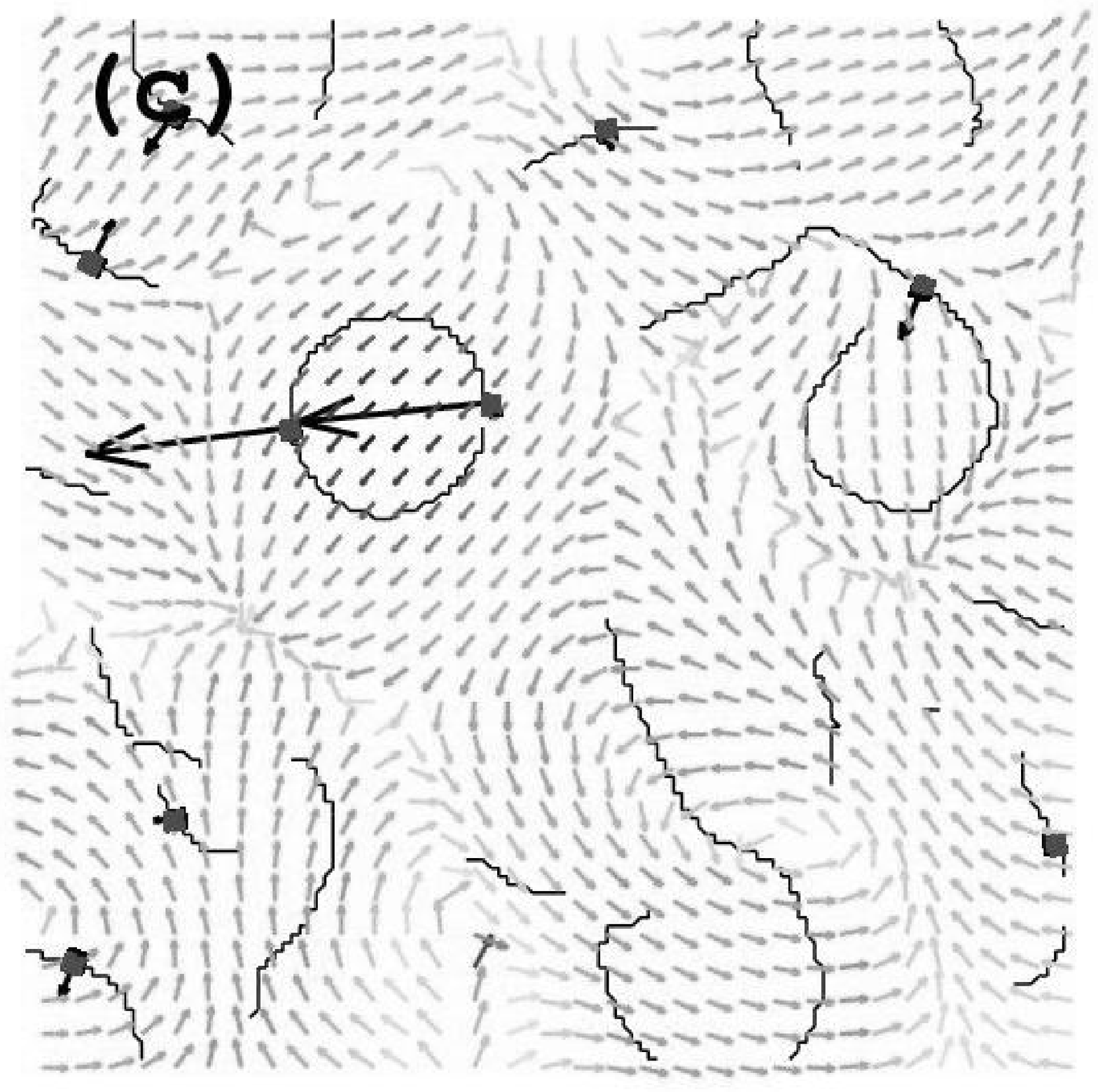}
  \end{minipage}
  \hfill
  \begin{minipage}[c]{.45\linewidth}
    \includegraphics[width=4.2cm]{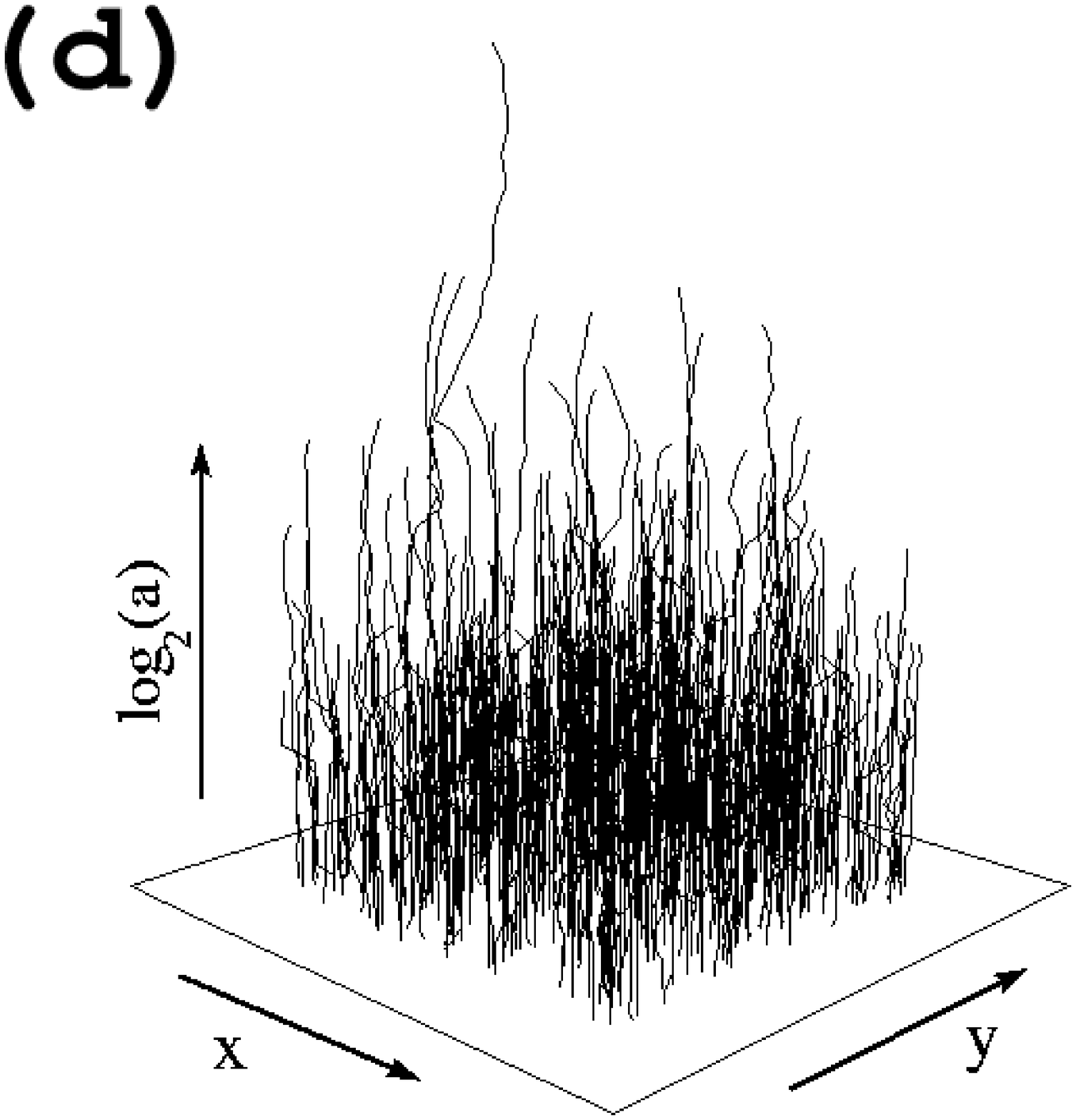}
  \end{minipage}
  \caption{
        \DD{} \WT{} analysis of the 2D vector-valued self-similar
         measure shown in Fig.~\ref{fig1} but with systematic random
         permutation of the $S_i$.
         $\bpsi$ is a first-order
        analyzing wavelet ($\phi(\gra{r})$ is the Gaussian).
        (a) 32 grey-scale coding of the central $(128)^2$ portion of the
         original $(1024)^2$ field.
         In (b) $a=2^2\sigma_W$ and (c) $a=2^3\sigma_W$, are shown the
         maxima chains; from the local maxima (\WTMMM{}) of $\Mpsi$ 
         along these chains ($\carreNoir$) originates a 
         black arrow whose length is proportional to $\Mpsi$ and direction
         is along $\Tpsirho[\mathbf{V}]$. 
         (d) \WT{} skeleton obtained by 
         linking the \WTMMM{} across scales. $\sigma_W=7$ (pixels) is
         the characteristic size of $\bpsi$ at the smallest resolved scale.
         \label{fig2}
  }
\end{figure}

\begin{figure}
  \centering
  \includegraphics[scale=0.48,angle=-90]{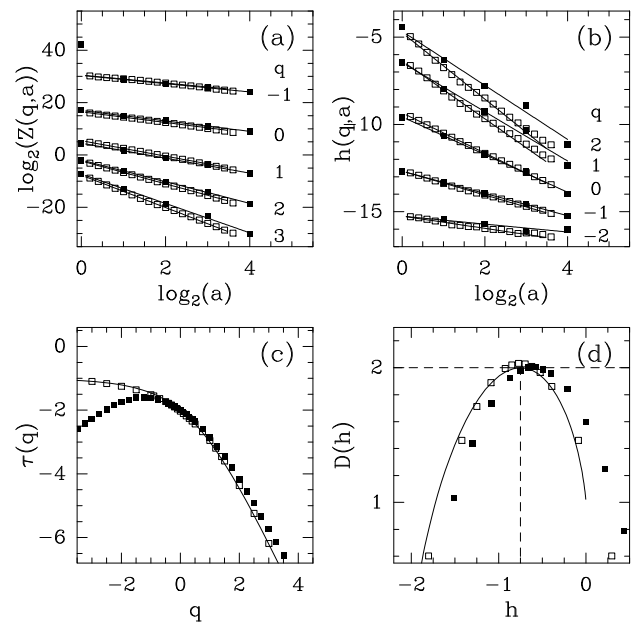}
  \caption{
        Multifractal analysis of the 2D vector-valued random
        measure field 
        using the \DD{} tensorial \WTMM{} method 
        ($\carreBlanc$) and BC techniques ($\carreNoir$).
        (a) $\log_2 {\cal Z}(q,a)$ vs $\log_2 a$; 
        (b) $h(q,a)$ vs $\log_2 a$;
        the solid lines correspond 
        to linear regression fits over
        $\sigma_W \lesssim a \lesssim 2^{4}\sigma_W$. 
        (c) $\tau(q)$ vs $q$; the solid line
        corresponds to the theoretical prediction (Eq.~(\ref{eq7})).
        (d) $D(h)$ vs $h$; 
        the solid line is the Legendre transform of Eq.~(\ref{eq7}).
        \label{fig3} 
      }
\end{figure}

As a test application of this extension of the WTMM method to the
vector situation, let us consider the self-similar 2D vector measures
supported by the unit square defined in Ref.~\cite{aFal96}.
As sketched in Fig.~\ref{fig1},
from
step $n$ to step $n+1$, each square is divided into 4 identical
sub-squares and for each of these sub-squares, one defines a
similitude $S_i$ that transforms the vector $\mathbf{V}^{(n)}$ at step
$n$ into the vector $\mathbf{V}_i^{(n+1)}$ for the sub-square $i$ at
step $n+1$. 
 The
$\sigma$-additivity property of positive scalar measures is now
replaced by 
the vectorial
additivity condition $\mathbf{V}^{(n)}=\sum_{i=1}^4
\mathbf{V}_i^{(n+1)}$. A straightforward calculation yields the
following analytical expression for the partition function scaling
exponents $\tau(q)$ (Eq.~(\ref{eq4})):
\addtocounter{equation}{1}
\begin{equation}
\tau(q) = -\log_2(p_1^q+p_2^q+p_3^q+p_4^q) -q,
\label{eq7}
\end{equation}
where $p_i$ ($i=1$ to 4), are the norms of the similitudes $S_i$. Note
that this formula is identical to the theoretical 
spectrum of a non-conservative scalar multinomial measure distributed
on the unit square with the weights
$p_i$~\cite{aKes03_prl,tKes03}.
Indeed, if the construction process in  Fig.~\ref{fig1} is
conservative from a vectorial point of view, it does not conserve the
norm of the measure: $\sum_{i=1}^4 p_i=4>1$. 
From Legendre
transforming Eq.~(\ref{eq7}), one gets a $D(h)$ singularity spectrum
with a characteristic multifractal single-humped shape (see Fig.~\ref{fig3}d)
supported by the interval 
$[h_{\min},h_{\max}]=[-1-\log_2(\max_i p_i),-1-\log_2(\min_i p_i)]$
and whose maximum $D_F=-\tau(0)=2$ is the signature that the considered
vector-valued measure 
is almost everywhere singular on the unit square.

In Fig.~\ref{fig2}
are illustrated the main steps of our tensorial WT methodology when
applied to 16 $(1024)^2$ realizations of a random generalization of the
vectorial multiplicative construction process described in
Fig.~\ref{fig1}. Focusing on the central $(128)^2$ sub-square, we show
the singular vector-valued measure (Fig.~\ref{fig2}a) and the
corresponding WTMM chains
computed with a first order analyzing wavelet at two different scales
(Figs~\ref{fig2}b and \ref{fig2}c).
On these maxima chains, the black dots correspond to the
location of the WTMMM at these scales. The size of the
arrows that originate from each black dot is
proportional to the spectral radius $\rho(\mathbf{b},a)$ and its
direction is along the
eigenvector $\mathbf{G}_{\rho}(\mathbf{b},a)$.
When linking these WTMMM across scales, one gets the set of maxima lines
shown in Fig.~\ref{fig2}d as defining the WT skeleton.
In Fig.~\ref{fig3}
are reported the results of the computation of the multifractal
spectra (annealed averaging).
As shown in Fig.~\ref{fig3}a, ${\mathcal Z}(q,a)$
(Eq.~(\ref{eq4})) display nice scaling behavior over four octaves (when
plotted versus $a$ in a logarithmic representation), for $q\in ]-2,4[$
for which statistical convergence turns out to be achieved. A linear
regression fit of the data yields the nonlinear $\tau(q)$ spectrum
shown in Fig.~\ref{fig3}c, in remarkable agreement with the theoretical
spectrum (Eq.~(\ref{eq7})). This multifractal diagnosis is confirmed
in Fig.~\ref{fig3}b where the slope of $h(q,a)$ (Eq.~(\ref{eq56}))
versus $\log_2 a$, clearly depends on $q$. From the estimate of $h(q)$
and $D(q)$ (Eq.~(\ref{eq56})), one gets the
single-humped $D(h)$ curve shown in Fig.~\ref{fig3}d which matches
perfectly the theoretical $D(h)$ spectrum.
In Fig.~\ref{fig3}, we have reported
for comparison, the results obtained when using a box-counting (BC)
algorithm adapted to the multifractal analysis of singular
vector-valued measures~\cite{tKes03,aFal96,remark2}. There is no doubt
that BC provides much poorer results, especially concerning the
estimates of $\tau(q)$, $h(q)$ and $D(q)$ for
negative $q$ values. This deficiency mainly results
from the fact that the vectorial resultant may be very small whereas
the norms of the vector measures in the sub-boxes are 
not small at all. The results reported in Fig.~\ref{fig3}
bring the demonstration that our tensorial WTMM methodology paves the
way from multifractal analysis of singular scalar measures to singular
vector measures.

\begin{figure}
  \centering
  \includegraphics[scale=0.48,angle=-90]{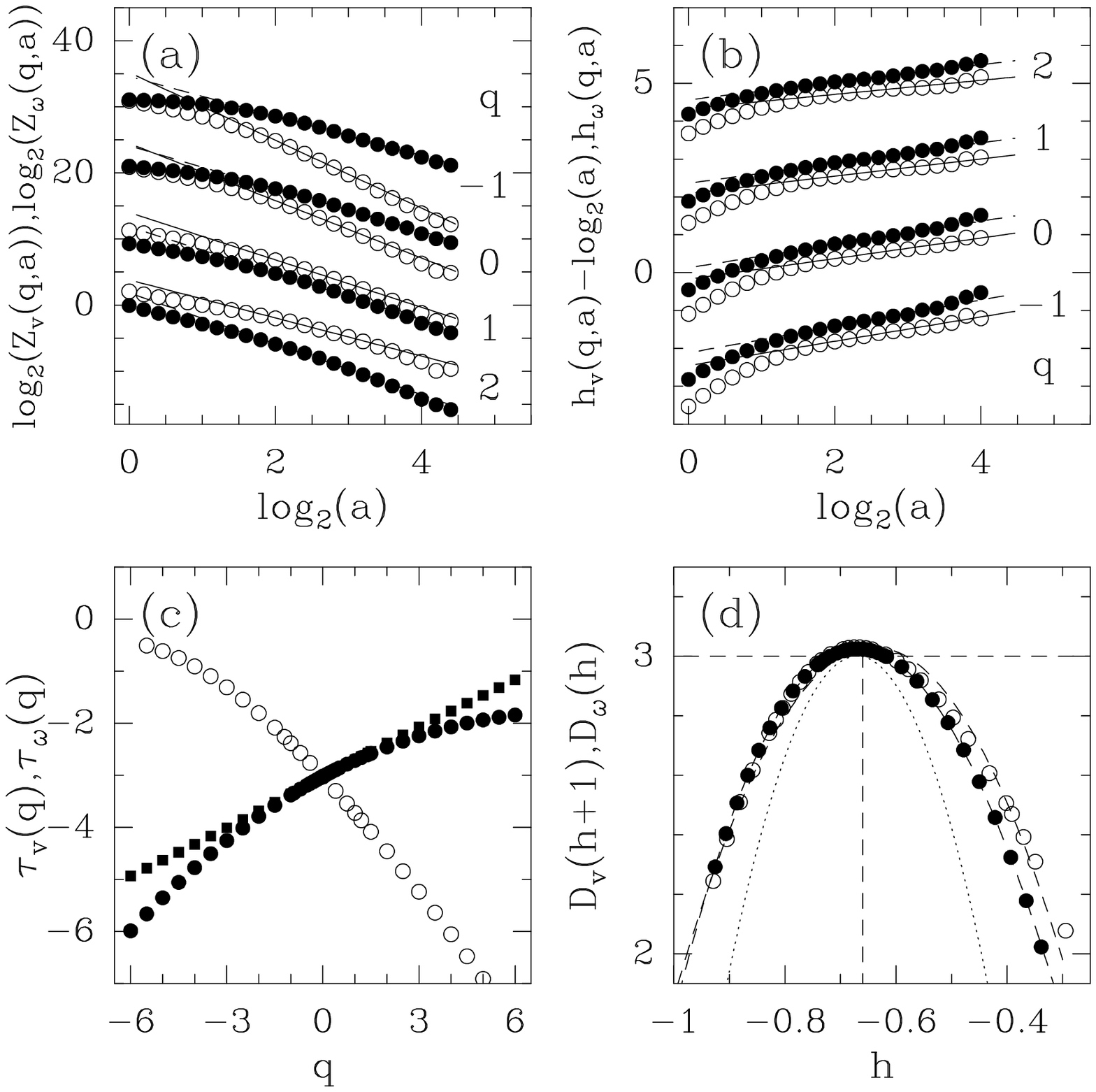}
  \caption{
        Multifractal analysis of L\'ev\^eque DNS velocity
        ($\rondNoir$) and vorticity ($\rondBlanc$) 
        fields ($d=3$, 18 snapshots)
        using the tensorial \TD{} \WTMM{} method;
        the symbols ($\carreNoir$) correspond to a similar analysis of
        vector-valued fractional Brownian motions, $\mathbf{B}^{H=1/3}$.
        (a) $\log_2 {\cal Z}(q,a)$ vs $\log_2 a$; 
        (b) $h_{\boldsymbol{\omega}}(q,a)$ vs $\log_2 a$ and
        $h_{\mathbf{v}}(q,a)-\log_2a$ vs $\log_2 a$;
        the solid and dashed lines correspond 
        to linear regression fits over
        $2^{1.5}\sigma_W \lesssim a \lesssim 2^{3.9}\sigma_W$.
        (c) $\tau_{\mathbf{v}}(q)$,  
        $\tau_{\boldsymbol{\omega}}(q)$ and
        $\tau_{\mathbf{B}^{1/3}}(q)$ vs $q$; 
        (d) $D_{\mathbf{v}}(h+1)$, 
        $D_{\boldsymbol{\omega}}(h)$ vs $h$; 
        the dashed lines correspond to log-normal regression fits with
        the parameter values 
        $C_2^{\mathbf{v}}=0.049$ and
        $C_2^{\boldsymbol{\omega}}=0.055$; the dotted line is
        the experimental singularity spectrum
        ($C_2^{\delta v_{\!/\!\!/}}=0.025$) for 1D
        longitudinal velocity increments~\cite{aArn98aMal00aDelArn01}. 
        \label{fig4}
      }
\end{figure}

In Fig.~\ref{fig4}
 are reported the results of the application of our tensorial WTMM
 method to isotropic turbulence DNS data obtained by L\'ev\^eque. This
 comparative 3D multifractal analysis of the velocity ($\mathbf{v}$)
 and vorticity ($\boldsymbol{\omega}$) fields corresponds to some
 averaging over 18 snapshots of $(256)^3$ DNS run at
 $R_\lambda=140$. As shown in Figs.~\ref{fig4}a and \ref{fig4}b, both
 the $\mathcal{Z}(q,a)$ and $h(q,a)$ partition functions display
 rather nice scaling properties for $q=-4$ to 6, except at small
 scales ($a\lesssim 2^{1.5}\sigma_W$) where some curvature is observed
 in the log-log plots likely induced by dissipation
 effects~\cite{bFri95,aArn98aMal00aDelArn01}. 
Linear regression fit of the data
 (Fig.~\ref{fig4}a) in the range $2^{1.5}\sigma_W \leq a \leq
 2^{3.9}\sigma_W$ yields the nonlinear $\tau_{\mathbf{v}}(q)$ and
 $\tau_{\boldsymbol{\omega}}(q)$ spectra shown in Fig.~\ref{fig4}c, the
 hallmark of multifractality. For the vorticity field,
 $\tau_{\boldsymbol{\omega}}(q)$ is a decreasing function similar to
 the one obtained for the singular vector-valued measure in
 Fig.~\ref{fig3}c; hence $h(q)$($=\partial \tau(q)/\partial q$)$<0$ and
 the support of the $D(h)$ singularity spectrum expands over negative
 $h$ values as shown in Fig.~\ref{fig4}d. In contrast
 $\tau_{\mathbf{v}}(q)$ is an increasing function 
 which implies that $h(q)>0$ as the signature that $\mathbf{v}$ is a
 continuous function. Let us point out that the so-obtained
 $\tau_{\mathbf{v}}(q)$ curve significantly departs from the linear
 behavior obtained for 18 $(256)^3$ realizations of vector-valued
 fractional Brownian motions $\mathbf{B}^{1/3}$ of index $H=1/3$, in
 good agreement with the theoretical spectrum
 $\tau_{\mathbf{B}^{1/3}}(q)=q/3-3$. But even more remarkable, the
 results reported in Fig.~\ref{fig4}b for $h(q,a)$ 
 suggest, up to statistical uncertainty, the validity of
 the relationship
 $h_{\boldsymbol{\omega}}(q)=h_{\mathbf{v}}(q)-1$. Actually, as shown
 in Fig.~\ref{fig4}d,
 $D_{\boldsymbol{\omega}}(h)$ and $D_{\mathbf{v}}(h)$ curves are likely to
 coincide after translating the later by one unit on the left.
This is to our knowledge the first numerical evidence that the
 singularity spectra of $\mathbf{v}$ and $\boldsymbol{\omega}$ might be
 so intimately related: $D_{\mathbf{v}}(h+1)=D_{\boldsymbol{\omega}}(h)$
 (a result that could have been guessed intuitively by noticing that
 $\boldsymbol{\omega}=\boldsymbol{\nabla}\wedge\mathbf{v}$ involves
 first order derivatives only).
Finally, let us note that, for both fields, the $\tau(q)$ and $D(h)$
 data are quite well fitted by log-normal parabolic
 spectra~\cite{aArn98aMal00aDelArn01}:
\begin{equation}
  \begin{aligned}
    \tau(q)&=-C_0-C_1q-C_2q^2/2\, ,\\
     D(h)&=\,C_0-(h+C_1)^2/2C_2\,.
    \label{eq8}
  \end{aligned}
\end{equation}
Both fields are found singular almost everywhere:
$C_0^{\mathbf{v}}=-\tau_{\mathbf{v}}(q=0)=D_{\mathbf{v}}(q=0)=3.02 \pm
0.02$ and 
$C_0^{\boldsymbol{\omega}}=3.01 \pm 0.02$.
The most frequent H\"older exponent $h(q=0)=-C_1$ (corresponding to
the maximum of $D(h)$) takes the value $-C_1^{\mathbf{v}}\simeq
-C_1^{\boldsymbol{\omega}}+1=0.34\pm0.02$. Indeed, this estimate is
much closer to the K41 prediction $h=1/3$~\cite{bFri95} than previous
experimental measurements ($h = 0.39\pm0.02$) based on the
analysis of longitudinal velocity
fluctuations~\cite{aArn98aMal00aDelArn01}. 
Consistent estimates are obtained for $C_2$ (that characterizes the width of
$D(h)$):
$C_2^{\mathbf{v}}=0.049\pm0.003$ and
$C_2^{\boldsymbol{\omega}}=0.055\pm0.004$.
Note that these values are much larger than the experimental estimate
$C_2=0.025\pm0.003$ derived for 1D longitudinal velocity
increment statistics~\cite{aArn98aMal00aDelArn01}.
Actually they are comparable to the value
$C_2=0.040$ extracted from experimental transverse
velocity increments~[19b].

To conclude, we have generalized the WTMM method to vector-valued
random fields. Preliminary applications to DNS turbulence data have
revealed the existence of an intimate relationship between the
velocity and vorticity 3D statistics that turn out to be significantly
more intermittent than previously estimated from
1D longitudinal velocity increments statistics.
This new methodology looks very promising to many extents. Thanks to
the SVD, one can focus on fluctuations that
are locally confined in 2D ($\min_i \sigma_i=0$) or in 1D (the two
smallest $\sigma_i$ are zero) and then simultaneously proceed to a
multifractal and structural analysis of turbulent flows.
The investigation along this line of vorticity sheets and vorticity
filaments in DNS is in current progress.
We are very grateful to E. L\'ev\^eque for allowing us to have access
to his DNS data and to the CNRS under GDR turbulence.


\begin{thebibliography}{10}
\bibitem{remark}
Note that if $h_i(\mathbf{r}_0)$ are the H\"older exponents
of the $d$ scalar components $V_i(\mathbf{r}_0)$ of $\mathbf{V}$, then
$h(\mathbf{r}_0)=\min_i h_i(\mathbf{r}_0)$.
\bibitem{remark2}
We recall the relationship $h=\alpha-d$, between the H\"older exponent
$h$ and the singularity exponent $\alpha$ generally obtained by BC
techniques ($D(\alpha-d)=f(\alpha)$).
\end{thebibliography}
\end{document}